\documentclass[10pt,letterpaper]{article}
\usepackage{geometry}
\usepackage{epsfig}
\usepackage{amsmath,amssymb,slashed,bbm}
\usepackage{setspace}
\csname@addtoreset\endcsname{equation}{section}

\makeatletter
\pdfoutput=1

\usepackage[T1]{fontenc}
\usepackage{ae}
\usepackage{aecompl}

\usepackage[english]{babel}

\usepackage[sf,bf,small]{titlesec}

\usepackage[small,sf,bf,hang]{caption}

\usepackage[multiple]{footmisc}
\long\def\symbolfootnote[#1]#2{\begingroup%
\def\thefootnote{\fnsymbol{footnote}}\footnote[#1]{#2}\endgroup}

\usepackage{titletoc}
\dottedcontents{section}[6mm]{\smallskip }{6mm}{0pc} 
\dottedcontents{subsection}[8mm]{\itshape\small }{6mm}{0pc}
\dottedcontents{subsubsection}[16mm]{\itshape\small }{10mm}{0pc}
\def\tableofcontents{\subsection*{\contentsname}\vspace{-2mm}\@starttoc{toc}}

\usepackage[nosort]{cite} 

\usepackage[parsep]{collref}

\usepackage{feynmp}  
\DeclareGraphicsExtensions{.pdf}  

\renewcommand{\vec}[1]{{\bf #1}}
\renewcommand{\bar}[1]{\overline{#1}}

\def \bea  {\begin{eqnarray}}
\def \eea  {\end{eqnarray}}

\newcommand{\bra}[1]{\langle{#1}|}
\newcommand{\ket}[1]{|{#1}\rangle}

\newcommand{\nn}{\nonumber}

\newcommand{\ns} \normalsize

\newcommand{\5}{$AdS_5\times S^5$}
\newcommand{\3}{$AdS_3\times S^3\times S^3\times S^1$}
\newcommand{\2}{$AdS_2\times S^2\times T^6$}

\AtBeginDocument{
  
}

\makeatother

\begin{document}
\begin{flushright}
MIFPA-12-33\\
QGaSLAB-12-04\bigskip\bigskip
\par\end{flushright}

\begin{center}
\textsf{\textbf{\Large Classical and quantum integrability in $AdS_2 / CFT_1$ \smallskip\smallskip }}\\
\textsf{\textbf{\Large  }}
\par\end{center}{\Large \par}

\begin{singlespace}
\begin{center}
Jeff Murugan$^{1}$, Per Sundin$^{1}$ and  Linus Wulff$^{2}$ \bigskip \\

{\small $^{1}$}\emph{\small{} The Laboratory for Quantum Gravity \& Strings}\\
\emph{\small Department of Mathematics and Applied Mathematics, }\\
\emph{\small University of Cape Town,}\\
\emph{\small Private Bag, Rondebosch, 7700, South Africa}{\small }\\
\emph{\small jeff@nassp.uct.ac.za, nidnus.rep@gmail.com}\vspace{0.2cm}

{\small $^{2}$}\emph{\small{} George P. \& Cynthia Woods Mitchell Institute for Fundamental Physics and Astronomy,}\\
\emph{\small Texas A\&M University, College Station, }\\
\emph{\small TX 77843, USA}\\
\emph{\small linus@physics.tamu.edu }{\small \bigskip }\emph{\small }\\
\par\end{center}{\small \par}
\end{singlespace}

\subsection*{\hspace{9mm}Abstract}
\begin{quote}
We investigate the type IIA string on $AdS_2\times S^2\times T^6$ supported by RR-flux which describes the gravitational side of the $AdS_2 / CFT_1$ correspondence. While the four-dimensional part $AdS_2\times S^2$ can be realized as a supercoset, the full superstring has both coset and non-coset excitations, the latter giving rise to massless worldsheet modes, a somewhat novel feature in $AdS / CFT$. The string is nevertheless known to be integrable at the classical level. In this paper we perform several computations checking aspects of both classical and quantum string integrability. At the classical level we compute energies for the near BMN string and successfully match these against Bethe ansatz predictions. Furthermore, integrability dictates a magnon dispersion relation which we compare with the poles of loop corrected propagators, at both the one and two-loop level. At one loop, where only tadpole diagrams contribute, we find that the bosonic and fermionic contributions sum up to zero. Under the assumption of worldsheet supersymmetry, we then compute the two-loop sunset diagram in the near flat space limit. As in \5 we find that the result fits nicely into the sine-square structure of the dispersion relation.
\bigskip  \thispagestyle{empty} 
\end{quote}

\newsavebox{\feynmanrules}
\sbox{\feynmanrules}{
\begin{fmffile}{diagrams} 


\fmfset{thin}{0.6pt}  
\fmfset{dash_len}{4pt}
\fmfset{dot_size}{1thick}
\fmfset{arrow_len}{6pt} 


\begin{fmfgraph*}(60,60)
\fmfkeep{sunset}
\fmfleft{i}
\fmfright{o}
\fmf{plain,tension=5}{i,v1}
\fmf{plain,tension=5}{v2,o}
\fmf{plain,left,tension=0.4}{v1,v2,v1}
\fmf{plain}{v1,v2}
\fmfdot{v1,v2}
\end{fmfgraph*}

\begin{fmfgraph*}(82,45)
\fmfkeep{tadpole-boson}
\fmfleft{in,p1}
\fmfright{out,p2}
\fmfdot{c}
\fmf{plain_arrow}{in,c}
\fmf{plain_arrow}{c,out}
\fmf{plain_arrow,right, tension=0.8, label=\small{$x_m$}}{c,c}
\fmf{phantom, tension=0.2}{p1,p2}
\end{fmfgraph*}

\begin{fmfgraph*}(82,45)
\fmfkeep{tadpole-fermion}
\fmfleft{in,p1}
\fmfright{out,p2}
\fmfdot{c}
\fmf{plain_arrow}{in,c}
\fmf{plain_arrow}{c,out}
\fmf{plain_arrow,right, tension=0.8, label=\small{$\chi_\pm^c$}}{c,c}
\fmf{phantom, tension=0.2}{p1,p2}
\end{fmfgraph*}

\end{fmffile}
}

\newpage
\setcounter{page}{1}
\tableofcontents{}

\section{Introduction}
$AdS / CFT$ dualities are arguably one of the most fascinating areas in contemporary theoretical physics. While the canonical example is $AdS_5 / CFT_4$, there are by now several other incarnations \cite{Maldacena:1997re,Gubser:1998bc,Witten:1998qj,Aharony:1999ti,Aharony:2008ug}. In this paper we will study the gravitational side of the $AdS_2 / CFT_1$ correspondence where the $CFT_1$ is, as of yet, perhaps the most illusive example of a boundary CFT  \cite{Maldacena:1997re}. It might be realized as a large $N$ superconformal quantum-mechanical system or a chiral two-dimensional CFT and is not very well understood \cite{Maldacena:1997re,Gibbons:1998fa,Strominger:1998yg,Maldacena:1998uz}. The dual string theory on the other hand is more accessible and is either a type IIA or IIB theory on \2 preserving eight supersymmetries. This geometry can be supported by different choices of RR-flux. We will consider a type IIA example with $F_2$ and $F_4$ flux, while the other type IIA/B examples can be obtained by performing T-dualities in the toroidal directions. Furthermore, the $AdS_2\times S^2$ factor is interesting in its own right since it appears as the near horizon limit of Reissner-Nordstr\"om black holes \cite{Maldacena:1997re}. 

The flat toroidal directions are a somewhat novel feature for $AdS / CFT$ and give rise to complications in the exact formulation based on the Bethe ansatz. The four-dimensional $AdS_2\times S^2$ factor can be described by the supercoset $PSU(1,1|2)/SO(1,1)\times SO(2)$ and one can try to realize the string as a supercoset sigma model \cite{Zhou:1999sm}, similar to $AdS_5\times S^5$ \cite{Metsaev:1998it} and $AdS_4\times \mathbbm{CP}^3$ \cite{Arutyunov:2008if,Stefanski:2008ik} (for a modified version of this approach see \cite{Berkovits:1999zq}). From the worldsheet perspective the excitations of the coset part enter as massive modes in the BMN limit and their spectrum should be completely encoded in sets of quantum Bethe equations \cite{Sorokin:2011rr}. These equations are the quantum counterpart of a set of classical finite gap equations which arise naturally from the equations of motion for the coset variables \cite{Zarembo:2010yz, Kazakov:2004qf,Beisert:2005bm}. Thus, for the supercoset part, the machinery of integrability, developed over the last decade in earlier incarnations of $AdS / CFT$, is applicable \cite{Beisert:2010jr}. However, the situation is not as simple as in other versions of $AdS / CFT$. First of all the flat directions are {\it not} part of the supercoset construction (although they can be added by hand) and give rise to massless worldsheet excitations which, naturally, are not captured by the finite gap technique. Second, and even more severe, is the fact that the supercoset contains only eight fermions, corresponding to the eight supersymmetries preserved by the background. This means that unlike in previously studied examples of $AdS / CFT$ \cite{Gomis:2008jt,Babichenko:2009dk,Rughoonauth:2012qd} the Green-Schwarz (GS) superstring in \2 can never be reduced to the supercoset model by gauge-fixing kappa symmetry, since this process necessarily leaves 16 physical fermions\footnote{The supercoset model turns out instead to be a consistent truncation (at the classical level) of the full GS string \cite{Sorokin:2011rr}.}. One therefore has to work with the full GS action. In fact in the full GS string action the coset and non-coset sectors do not decouple beyond leading order but are mixed through interaction terms involving the fermions \cite{Sorokin:2011rr}. For the purely bosonic string, the two sectors can be added as a linear sum (modulo Virasoro constraints), but once the fermionic directions are taken into account there is non-trivial mixing. Consequently, the physical spectrum involves both coset and non-coset modes. Nevertheless, even though the two sectors do not decouple, there are still reasons to believe that the full model is integrable. The first step toward a proof of this was presented in \cite{Sorokin:2011rr}, where the authors demonstrated classical integrability (up to quadratic order in fermions) of the GS string action (see also \cite{Cagnazzo:2011at} where this was done to quadratic order in the non-coset fermions). Furthermore, in the same paper a set of asymptotic Bethe equations were also presented. However, as stated above, the massless non-coset modes remain somewhat mysterious and were not incorporated in the Bethe equations. While they are believed to enter as intermediate states (internal propagators in Feynman diagrams), incorporating them as external states remains an open problem. For a recent discussion see \cite{OhlssonSax:2011ms}.

In this paper we provide further evidence for integrability of the \2 string. We start out by writing the GS Lagrangian (to quadratic order in fermions) to quartic order in fields utilizing a BMN-like expansion \cite{Berenstein:2002jq}. Equipped with the quartic Lagrangian we perform several computations both at tree-level, one-loop and finally two-loop level. For the classical analysis we compare the near-BMN energy corrections for the bosons with the predictions of the conjectured Bethe equations, similar to the analysis performed in \cite{Frolov:2006cc,Hentschel:2007xn,Astolfi:2008ji,Sundin:2008vt,Astolfi:2011ju,Astolfi:2011bg}. As expected we find complete agreement. We then investigate quantum corrections to the magnon dispersion relation
\bea
\label{eq:integrability-energy}
E=\sqrt{1+4h(g)^2\sin^2\frac{p_1}{2}},\qquad h(g)=\frac{g}{2\pi}+\dots\,,
\eea 
which is fixed by integrability up to the unknown function $h(g)$ \cite{Sorokin:2011rr,Nishioka:2008gz,Gaiotto:2008cg,Grignani:2008is,Gromov:2008fy}. This dispersion relation should coincide with the loop corrected pole of propagators for the massive string modes making it possible to compare the magnon energies with explicit string theory computations \cite{Klose:2007rz, Abbott:2011xp, Sundin:2012gc}. At one loop we find that the correction for the massless modes is trivially zero while the one-loop correction for the massive modes is zero due to delicate cancellations between boson and fermion loops, as happens also in \5 and $AdS_4 \times \mathbbm{CP}^3$. It follows then, that there is no subleading one-loop term in $h(g)$ and the first correction to the dispersion relation enters at two loops\footnote{The regularization ambiguities present in \3 and $AdS_4 \times \mathbbm{CP}^3$ do not occur since there is no notion of composite modes here, see \cite{Gromov:2008fy,Sundin:2012gc,McLoughlin:2008ms, Alday:2008ut,Krishnan:2008zs,Shenderovich:2008bs,McLoughlin:2008he,Abbott:2010yb, Zarembo:2009au, LopezArcos:2012gb}.}. A full-blown two-loop computation generally demands the Lagrangian up to sixth order in fields, something which is not available at present. However, utilizing the near flat space (NFS) limit, the two-loop contribution is in fact determined solely by the quartic Lagrangian \cite{Maldacena:2006rv,Klose:2007rz}. Furthermore, under the assumption of worldsheet supersymmetry we can bypass the need for the quartic fermion terms. Thus, the NFS expanded dispersion relation is completely determined by a four-vertex sunset diagram. We compute this diagram explicitly and show that the result fits nicely with the sine-square structure of the dispersion relation. Thus, we are able to demonstrate two-loop quantum integrability for the \2 string (in the NFS limit). We furthermore observe that the sunset diagram is a sum of diagrams with both massive and massless internal propagators, showing explicitly that the massless modes contribute to the amplitude as virtual particles, similarly to what happens for the \3 string \cite{Sundin:2012gc}. In fact the specific interactions of the non-coset fermions present in the GS action appear to be crucial for getting a result compatible with integrability. This provides some evidence for the integrability of the full GS string action at the quantum level.

The outline of this article is as follows. Section 2 describes the GS string in \2 and its near BMN expansion to quartic order in fields (although only quadratic in fermions). In section 3 we compute string energies for some closed rank-one sectors and successfully match these against predictions of the Bethe equations proposed in \cite{Sorokin:2011rr}. In section 4 we then compute one-loop corrections to two-point functions and demonstrate that these are either identically zero or sum up to zero. And finally the two-loop computation in the NFS limit is done where, after a rather lengthy computation, we end up with an expression that fits nicely into the sine-square dispersion relation. We conclude the paper with a discussion and appendices describing our choice of coordinates. 

\section{Green-Schwarz superstring in \2}
\label{stringsection}
The starting point for the analysis of this paper is the Green-Schwarz superstring action in \2. Restricting to quadratic order in fermions, both the type IIA and IIB GS string action is known in closed form \cite{hep-th/9601109,hep-th/9907202}. For simplicity we will consider only the type IIA case here since in this case we can combine the two Majorana-Weyl spinors of the type IIA superspace into a single 32-component Majorana spinor, simplifying somewhat our analysis. The type IIB  \2 solutions and different type IIA solutions with RR-flux are related to each other by T-dualities along the toroidal directions (see for example \cite{Sorokin:2011rr}). We will therefore consider only one of these type IIA \2 solutions here.

\subsection{GS superstring to quadratic order in fermions in general IIA background}
The action for the GS superstring in a type IIA supergravity background
(with no NS--NS flux, and constant dilaton $\phi_0$) takes the following form up to quadratic
order in fermions \cite{hep-th/9601109,hep-th/9907202}
\begin{equation}
\label{action}
S = \frac{g}{2}\int\left(\frac{1}{2}\ast e^Ae_A+i*e^A\,\bar{\Theta}\Gamma_A{\mathcal D}\Theta-ie^A\,\bar{\Theta}\Gamma_A\Gamma_{11}{\mathcal D}\Theta\right)\,,
\quad\mbox{where}\quad g\sim T\sim\frac{R^2}{\alpha'}\,.
\end{equation}
The $e^A(X)$ $(A=0,1,\cdots,9)$ are worldsheet pullbacks of the vielbein one-forms of the purely bosonic part of the background, $*$ denotes the worldsheet Hodge-dual, and the generalized covariant derivative acting on the fermions is given by
\begin{equation}
\label{E}
{\mathcal D}\Theta=(\nabla-\frac{1}{8}e^A\,\slashed F\Gamma_A)\ \Theta\quad\mbox{where}\quad \nabla\Theta=(d-\frac{1}{4}\omega^{AB}\Gamma_{AB})\Theta\,,
\end{equation}
where $\omega^{AB}$ is the spin connection of the background space-time. The coupling to the RR fields comes in through the matrix
\begin{equation}
\label{eq:slashedF}
\slashed F=e^{\phi_0}\left(-\frac{1}{2}\Gamma^{AB}\Gamma_{11}F_{AB}+\frac{1}{4!}\Gamma^{ABCD}F_{ABCD}\right)\,.
\end{equation}
The 32-component spinor $\Theta$ satisfies the Majorana condition
\bea 
\label{spinor-conjugation}
\bar{\Theta}=\Theta^\dagger \Gamma_0=\Theta^t\mathcal C\,,
\eea 
where $\mathcal C$ is the charge conjugation matrix. We now turn to the specific background of interest here, \2.

\subsection{GS string in type IIA \2 and BMN expansion}
There are several different type II \2 supergravity solutions supported by RR-flux and preserving eight supersymmetries. They can be realized as the near horizon geometry of brane intersections \cite{Klebanov:1996mh,Sorokin:2011rr} and are related to each other by T-dualities along the toroidal directions. Here we will focus on one particular type IIA solution with non-zero $F_2$-flux through $AdS_2$ and non-zero $F_4$-flux through a combination of $S^2$ and $T^6$, 
\bea 
&& F_2=-\frac{e^{-\phi_0}}{2}e^b e^a \varepsilon_{ab}, \\ \nn 
&& F_4=-\frac{e^{-\phi_0}}{2}e^{\hat b} e^{\hat a} \varepsilon_{\hat a\hat b} J\,,
\eea 
where $a,b=0,1$, $\hat a,\hat b=2,3$, $\varepsilon^{01}=1=\varepsilon^{23}$ and $J$ is the K\"ahler form on $T^6$ which we take to be 
\begin{equation}
J=-dx^4dx^5-dx^6dx^7-dx^8dx^9\,.
\end{equation}
Note that the fluxes break the local $SO(6)$ invariance of the $T^6$ space to $U(3)$ via this choice of K\"ahler form. Using these expressions in (\ref{eq:slashedF}) we get
\bea 
\label{eq:F-slash}
\slashed F=-4\mathcal{P}_8\Gamma^{01}\Gamma_{11}\,,
\eea 
where $\mathcal{P}_8$ is a projector that projects on the eight supersymmetries preserved by the background. Explicitly it is given by
\bea 
\label{eq:P-projector}
\mathcal{P}_8=\frac{1}{8}\big(2-i\slashed J\gamma_7\big)\,,\qquad \slashed J=J_{a'b'}\Gamma^{a'b'}=2\big(\Gamma^4\Gamma^5+\Gamma^6\Gamma^7+\Gamma^8\Gamma^9\big)\,,\qquad\gamma_7=i\Gamma^4\cdots\Gamma^9\,.
\eea 
Note that this projector commutes with the gamma-matrices of the $AdS_2\times S^2$ part of the background, $[\mathcal{P}_8,\Gamma_a]=[\mathcal{P}_8,\Gamma_{\hat a}]=0$. The metric and spin connection for \2 is given in appendix A.

We are interested in the near BMN expansion for a string moving along the $\varphi$-direction of $S^2$ close to the speed of light (see appendix A for our choice of coordinates). In order to remove the unphysical fermionic degrees of freedom we will use a standard light-cone kappa symmetry gauge-fixing adapted to the BMN limit
\bea 
\label{eq:gauges}
\Gamma^+\Theta=0,\qquad \Gamma^\pm=\frac{1}{2}(\Gamma^0\pm\Gamma^3)\,.
\eea 
Furthermore, introducing light-cone coordinates as $x^\pm=\frac{1}{2}(t\pm\varphi)$ the action (\ref{action}) becomes, to leading order in the BMN expansion 
\bea 
\label{eq:L2ferm}
\frac{2}{g}\mathcal{L}_2&=&\frac{1}{2}\gamma^{ij}\partial_i x^m \partial_j x_m - \frac{1}{2}\gamma^{ij} \partial_i x^+\partial_j x^+ (x_1^2+x_2^2\big)
\nn \\ 
&&{}
-2i\,\bar\Theta\big(\gamma^{ij}+\varepsilon^{ij}\Gamma_{11}\big)\partial_i x^+\Gamma^-\partial_j\Theta
-2i\gamma^{ij}\partial_ix^+\partial_jx^+\,\bar{\Theta}\Gamma^-\Gamma^1\Gamma_{11}\mathcal P_8\Theta\,,
\eea 
where $\gamma^{ij}=\sqrt{-h}h^{ij}$ is the conformal worldsheet metric, the index $m$ runs over all eight transverse directions and all coordinates are dimensionless. It is clear from this expression that only the eight supercoset fermions (four after kappa symmetry gauge-fixing), which satisfy $\Theta=\mathcal P_8\Theta$, get a non-zero mass while the other fermions remain massless.

The above is actually the BMN string prior to light-cone gauge fixing. In detail the limit is specified by scaling the transverse degrees of freedom as
\bea \nn 
x_m \rightarrow \sqrt\frac{2}{g} x_m,\qquad \Theta\rightarrow \sqrt\frac{2}{g}\Theta,\qquad g\rightarrow \infty\,.
\eea 
The bosonic worldsheet parameterization invariance is fixed by \cite{Arutyunov:2005hd, Frolov:2006cc}
\bea
\label{eq:gauge}
x^+=\tau,\qquad p_-=\frac{\delta \mathcal{L}}{\delta \dot{x}^-}=1\,.
\eea
To lowest order this gives $\gamma^{ij}=\eta^{ij}$ and the quadratic Lagrangian becomes (the worldsheet metric has signature $(+-)$ in our conventions)
\bea 
\mathcal{L}_2&=&
\frac{1}{2}\big(\partial_+x^m\partial_-x_m -x_1^2-x_2^2\big)
-2i\,\bar\Theta_+\Gamma^-\partial_-\Theta_+
-2i\,\bar\Theta_-\Gamma^-\partial_+\Theta_-
+4i\,\bar{\Theta}_+\Gamma^-\Gamma^1\mathcal P_8\Theta_-\,,
\nn \\ 
\eea 
where $\partial_\pm=\partial_0\pm\partial_1$ and $\Theta_\pm=\frac{1}{2}(1\pm\Gamma_{11})\Theta$ are left/right-moving spinors.
One can diagonalize the fermionic terms by a suitable choice of basis, see (\ref{eq:theta}) for the explicit form of $\Theta$. The quadratic Lagrangian above then becomes
\bea 
\label{eq:L2-gf}
\mathcal{L}_2=\frac{1}{2}\big(\partial_+ x^m \partial_- x_m -x_1^2-x_2^2\big)
+\frac{i}{2}\chi^c_+\partial_-\chi^c_+
+\frac{i}{2}\chi^c_-\partial_+\chi^c_-
-i\,\chi^1_-\chi^1_+
-i\,\chi^2_-\chi^2_+\,,
\eea 
where $c=1,\dots,8$. Thus we find the following BMN spectrum \cite{Sorokin:2011rr}
\bea \nn
&& \underline{m=1}:\qquad \textrm{Bosons:}\qquad x_1, x_2 \qquad \,\,\,\,\textrm{Fermions:}\qquad \chi^1_\pm, \chi^2_\pm \\ \nn 
&& \underline{m=0}:\qquad \textrm{Bosons:}\qquad x_k \qquad \qquad \textrm{Fermions:}\qquad \chi^k_\pm\qquad  \qquad k=3,\dots,8
\eea 
so we have two real massive bosonic and fermionic coordinates and six transverse bosonic and fermionic massless coordinates. The appearance of massless coordinates is a rather novel feature in examples of AdS/CFT, see \cite{Babichenko:2009dk,Sorokin:2011rr}.

Beyond the quadratic approximation the BMN expansion gives a series in inverse powers of the coupling
\bea 
\mathcal{L}=\mathcal{L}_2+\frac{1}{g}\mathcal{L}_4+\mathcal{O}(g^{-2})\,.
\eea 
Consistency of the gauge-fixing (\ref{eq:gauge}) at higher order in perturbation theory demands that we add sub-leading corrections to the worldsheet metric. The precise form of the corrections is found by looking at the equations of motion for $x^-$ and a quick calculation gives
\bea 
\gamma_{ij}=\eta_{ij}+\frac{2}{g}\hat\gamma_{ij},\qquad \hat\gamma_{00}=\hat\gamma_{11}=-\frac{1}{2}\big(x_1^2-x_2^2\big),\qquad \hat\gamma_{01}=0\,.
\eea 
Expanding the Lagrangian in (\ref{action}) to the next order we get
\bea 
\label{eq:quarticL}
\mathcal{L}_4&=&
\frac{1}{2}\big(\partial_+x_1\partial_- x_1\,x_1^2-\partial_+x_2\partial_-x_2 \,x_2^2\big)
-\frac{1}{4}\big((\partial_+ x_m)^2+(\partial_- x_m)^2\big)(x_1^2-x_2^2)
\nn\\
&&{}
+i(x_1^2-x_2^2)\,\bar{\Theta}_+\Gamma^-\partial_+\Theta_+
+i(x_1^2-x_2^2)\,\bar{\Theta}_-\Gamma^-\partial_-\Theta_-
\nn\\
&&{}
-2i\partial_-x^m\partial_+x^n\,\bar{\Theta}_-\Gamma_m\Gamma^-\Gamma^1\mathcal P_8\Gamma_n\Theta_+
+i\partial_+x^m\,\bar{\Theta}_+\Gamma^-\Gamma_m(x_1\Gamma^1-x_2\Gamma^2)\Theta_+
\nn\\
&&{}
+i\partial_-x^m\,\bar{\Theta}_-\Gamma^-\Gamma_m(x_1\Gamma^1-x_2\Gamma^2)\Theta_-
+\mathcal{O}(\Theta^4)\,,
\eea 
where the indices $m,n$ run over the 8 transverse directions. The last two terms come from the spin connection (see appendix A). We have chosen not to decompose $\Theta$ in terms of $\chi$ to keep down the number of terms. The above Lagrangian is our starting point for the computations described in the following sections.

\section{Bosonic energy shifts and Bethe equations}
In \cite{Sorokin:2011rr} a set of quantum Bethe equations for the massive excitations were derived. In this section we will investigate whether we can match the tree-level energy corrections for the bosonic fields in the near BMN limit to the predictions of the Bethe equations, for similar analysis in other contexts see \cite{Callan:2004uv,Callan:2004ev,McLoughlin:2004dh,Sundin:2009zu,Astolfi:2009qh,Frolov:2006cc,Hentschel:2007xn,Astolfi:2008ji,Sundin:2008vt,Astolfi:2011ju,Astolfi:2011bg,Grignani:2008is,Rughoonauth:2012qd}. In order to do this we need the bosonic terms in the string Hamiltonian. From the definition of the conjugate momenta we find, dropping the fermions,
\bea \nn 
\dot x_1=p_1-\frac{1}{2}p_1x_2^2,\qquad 
\dot x_2=p_2+\frac{1}{2}p_2x_1^2,\qquad 
\dot x_k=p_k+\frac{1}{2}p_k\big(x_1^2-x_2^2\big),\quad k=3,\dots,8\,.
\eea 
Using this in the Legendre transformation of the string Lagrangian we get the (bosonic) Hamiltonian
\bea 
\label{eq:hamiltonian}
&& \mathcal{H}_{lc}=-P_+\\ \nn 
&& =\frac{1}{2}\big((x'_m)^2+(p_m)^2+x_1^2+x_2^2\big)+\frac{1}{2g}\Big[\big((x'_m)^2 + (p_m)^2\big)(x_1^2-x_2^2)+\big((x'_1)^2+p_1^2\big)x_1^2-\big((x'_2)^2+p_2^2\big)x_2^2\Big]\,,
\eea 
where the index $m$ runs over all eight transverse directions. Following the analysis in \5 we will assume that the quartic Hamiltonian is normal ordered \cite{Callan:2004uv,Callan:2004ev}. In principle this is an assumption and it would be interesting to perform a more rigorous analysis utilizing the full supersymmetric Hamiltonian. We might return to this question in a future publication\footnote{For example, the quartic Hamiltonian of the \3 and $AdS_4 \times \mathbbm{CP}^3$ string is not normal ordered \cite{Rughoonauth:2012qd,Astolfi:2011ju,Astolfi:2011bg}.}.

The oscillator expansion of the fields that diagonalizes the massive part of the quadratic Hamiltonian is given by
\bea 
\label{eq:field-exp}
&& x_c=\frac{1}{\sqrt{2\pi}}\int dp \frac{1}{\sqrt{2\omega_p}}\big(a_c(p) e^{-i p\cdot \sigma}+a_c(p)^\dagger e^{i p \cdot\sigma}\big)\,,
\eea 
with $c=1,2$ and $\omega_p=\sqrt{1+p^2}$. 

In order to obtain the energy corrections from the quartic Hamiltonian we will consider in-states of the form
\bea \label{eq-instates}
\ket{1,p_A}=\prod_{i=1}^A a_1(p_i)^\dagger \ket{0},\qquad \ket{2,p_A}=\prod_{i=1}^A a_2(p_i)^\dagger \ket{0}\,.
\eea 
Using the notation $\mathcal{H}_{lc}=\mathcal{H}_2+\frac{1}{g}\mathcal{H}_4$, the energies of the states (\ref{eq-instates}) are computed from
\bea \nn 
-P_+=\bra{c,p_A}\big(\mathcal{H}_2+\frac{1}{g}\mathcal{H}_4\big)\ket{c,p_A},\qquad c=1,2
\eea 
which after some algebra gives
\bea \label{eq:string-energies}
-P_+=\sum_{i=1}^A \omega_{p_i}-(-1)^c\frac{1}{2g}\sum_{i\neq j}^{A} \frac{p_i^2+p_j^2}{\omega_{p_i}\omega_{p_j}}\,.
\eea 
This result is similar, but not identical, to the corresponding energies of the rank one sectors of the $AdS_5\times S^5$ and $AdS_3\times S^3\times T^4$ strings. There the quartic piece came with a $(p_i+p_j)^2$ numerator instead of the separate squares as above. This is a simple consequence of the fact that there is no conserved $U(1)$ charge for the transverse $AdS_2$ and $S^2$ directions.

Let's see if we can reproduce this result from the Bethe equations. In \cite{Sorokin:2011rr} the first hints of integrability for the \2 string were presented. As for the \3 string, the critical spectrum mixes the coset and non-coset part in a non-trivial way. Nevertheless one can use the algebraic properties of the coset part to write down a set of Bethe equations. These techniques fall under the finite gap method which in turn can be seen as a classical limit of a set of conjectured quantum Bethe equations. In \cite{Sorokin:2011rr} the full Bethe equations for the coset $\frac{PSU(1,1|2)}{SO(1,1)\times SO(2)}$ were presented. Here, however, we only need the part describing rank one excitations, 
\bea 
\label{eq:rank1BE}
\Big(\frac{x^+_k}{x^-_k}\Big)^L=\prod_{j\neq k}^A \frac{x^+_k-x^-_j}{x^-_k-x^+_j} \frac{1-\frac{1}{x^+_k x^-_j}}{1-\frac{1}{x^-_k x^+_j}} \sigma^4_{BES}(k,j)\,.
\eea 
The Zhukovsky variables are given by
\bea 
\label{eq:zhukovsky}
x^\pm(p_k)=\frac{e^{\pm i\frac{p_k}{2}}\csc \frac{p_k}{2}}{2h(g)}\sqrt{1+4h(g)^2\sin\frac{p_k}{2}}
\eea 
and $\sigma_{BES}$ is the BES / BHL phase \cite{Beisert:2006ib,Beisert:2006ez}.
Finally, the function $h(g)$ is not determined by integrability but in order for it to match the relativistic magnon dispersion relation we have
\bea \label{eq:h}
h(g)=\frac{g}{2\pi}\,.
\eea 
The length parameter in (\ref{eq:rank1BE}) relates to the coupling and excitation number as\footnote{To leading order $L=\frac{1}{2}P_-$ since $2g=p_-$, where $p_-$ is the conjugate worldsheet momentum density of $x^-$. In the light-cone gauge we employ, $p_-=1$.}
\bea 
\label{eq:lengthparameter}
L=g+A-E
\eea 
where 
\bea 
E=ih(g) \sum_k^A\big(\frac{1}{x^+_k}-\frac{1}{x^-_k}\big)=-A+\sum_k^A \sqrt{1+p_k^2}+\dots
\eea 
Assuming that the rapidity momenta $p_k$ have a BMN scaling as
\bea 
p_k=\frac{p^0_k}{2g}+\frac{p^1_k}{(2g)^2}+\dots
\eea 
together with the constraint
\bea 
\sum_k^A p_k=0
\eea 
one can with a bit of work indeed reproduce the string energies as computed in (\ref{eq:string-energies}). That the Hamiltonian analysis and Bethe equations give the same energy corrections is completely expected. The computation only involves the massive part of the spectrum, meaning that only the coset excitations contribute. This stands in contrast to the \3 case, where even the classical energies involved massless excitations as internal lines. 

\section{Quantum corrected dispersion relation}
So far we have considered essentially the classical string where everything should fit nicely into the Bethe ansatz of \cite{Sorokin:2011rr}. However, since the Bethe ansatz is derived from the coset part only, it's not clear at all whether it could describe the full critical string with massless modes included. In this section we aim to investigate this issue by computing loop corrections to the dispersion relation of the bosonic modes. The dispersion relation coincides with the pole of the two-point function, so the computation basically boils down to determining the loop corrected propagator. 
We will start out by calculating the leading loop correction for bosonic excitations utilizing the full BMN string. We then turn to an explicit two-loop computation. In principle this should not be possible since we lack both the six-vertex interactions and the quartic fermion terms. However, utilizing a certain limit, the so-called near flat space limit (NFS) \cite{Maldacena:2006rv}, one can actually obtain the two-loop corrections without knowing these terms \cite{Klose:2007rz}. Intriguingly, the loop corrections of the massive coordinates fits nicely into the sine expression (\ref{eq:integrability-energy}), similar to $AdS_5 / CFT_4$.\footnote{For some interesting new research utilizing the NFS string in \5 see \cite{Klose:2012ju}.}

It follows from our one and two loop computation that the unknown interpolating function $h(g)$ does not receive any corrections (at least to this order in perturbation theory).

\subsection{One loop} 
The function $h(g)$ in (\ref{eq:integrability-energy}) is not determined by integrability and it could very well receive corrections at loop level. In this section and the next we intend to investigate this issue. However, before that, let's recall what happened in other string backgrounds. In both $AdS_4\times \mathbbm{CP}^3$ and \3 it has been found that $h(g)$ receives corrections, at least in certain regularization schemes \cite{Gromov:2008fy, Sundin:2012gc,McLoughlin:2008ms, Alday:2008ut,Krishnan:2008zs,Shenderovich:2008bs, Zarembo:2009au,Forini:2012bb}. In both cases this can be traced back to the presence of heavy modes, which in the Bethe ansatz are treated as composite states of two lighter ones. Furthermore, the string Lagrangian has a cubic interaction term that can mediate these heavy to light-light decay processes \cite{Zarembo:2009au, Rughoonauth:2012qd}. However, for the \2 string, there are only modes of one mass and the Bethe ansatz does not indicate that they are of a composite nature. Furthermore, the next-to-leading order Lagrangian is quartic in fields and does not mediate any decay processes. For this reason we expect that there will be no non-trivial one-loop correction to the propagators. However, this needs to be checked by actual computations. If there are non zero corrections, then they could be described as a one-loop term $c$ with
\bea \nn 
h(g)=\frac{g}{2\pi}+c\,.
\eea 
This implies that at large $g$, the dispersion relation (\ref{eq:integrability-energy}) expands as
\bea
E=\sqrt{1+4 (\frac{g}{2\pi}+c)^2\sin^2\frac{p_1}{2}}=\sqrt{1+p^2}+\frac{2\pi c \,p^2}{g\sqrt{1+p^2}}+\mathcal{O}(g^{-2})\,.
\eea
If $c$ is non-zero, then the one-loop correction to the two-point functions is also non-zero. Conversely, by explicitly computing the diagrams
\begin{align*}
\mathcal{A}_{B}^i	 & =\parbox[top][0.8in][c]{1in}{\fmfreuse{tadpole-boson}}+\parbox[top][0.8in][c]{1.5in}{\fmfreuse{tadpole-fermion}}
\end{align*}
for each bosonic coordinate, we can probe the subleading terms in $h(g)$.

Starting out with the two massive coordinates, labeled by $a=1,2$, we find using (\ref{eq:quarticL}) 
\bea 
i\mathcal{A}_a(\vec p^2=1)=-i(-1)^a \Big\{\big[2I[1]^0_1\big]_F-\big[I[1]^0_1+I^1_1[1]\big]_B\Big\}\,,
\eea 
where the extra subscript denotes whether a boson or a fermion propagates in the loop. We've introduced the following short hand notation
\bea \nn
I^s_n[\Delta]=\int\frac{d^2k}{(2\pi)^2}\frac{(\vec k^2)^s}{(\vec k^2-\Delta)^n}
\eea 
for the loop integrals. Evaluating the integrals in dimensional regularization immediately gives
\bea 
i\mathcal{A}_a(\vec p^2=1)=\frac{(-1)^a}{2\pi}\Big\{\big[\gamma-\frac{2}{\epsilon}+\log\pi\big]_F-\big[\gamma-\frac{2}{\epsilon}+\log\pi\big]_B\Big\}=0\,.
\eea 
Thus, as expected, we see that the massive modes have a vanishing one-loop correction to their two-point functions. Furthermore, since there is no notion of heavy composite modes, as for $AdS_4\times \mathbbm{CP}^3$ and \3, there should be no regularization ambiguity. 

For the massless modes a quick calculation gives
\bea 
i\mathcal{A}_k(\vec p^2=0)=0,\qquad k=3,\dots,8
\eea 
thus the one-loop correction is zero also for the massless bosonic excitations. Here each integral is separately zero and there are no cancellations between boson and fermion loops.

We conclude that, as in $AdS_5\times S^5$, the magnon dispersion relation (\ref{eq:integrability-energy}) does not receive a one-loop correction. For two loops, however, we expect the situation to be different due to the sine structure. In the next section we analyze this in detail. 

\subsection{Two-loop dispersion relation in the NFS limit}
In \cite{Maldacena:2006rv} an interesting limit of the worldsheet sigma model was proposed. This limit, dubbed the near-flat space limit (NFS) or Maldacena-Swanson (MS) limit, is basically a BMN expansion augmented with a Lorentz boost of the right-moving sector of the worldsheet theory. The NFS string is considerably simpler than the near BMN string, but nevertheless maintains some non-trivial features. Technically the limit is defined by
\begin{figure}[t]
\centering
\includegraphics[scale=0.55]{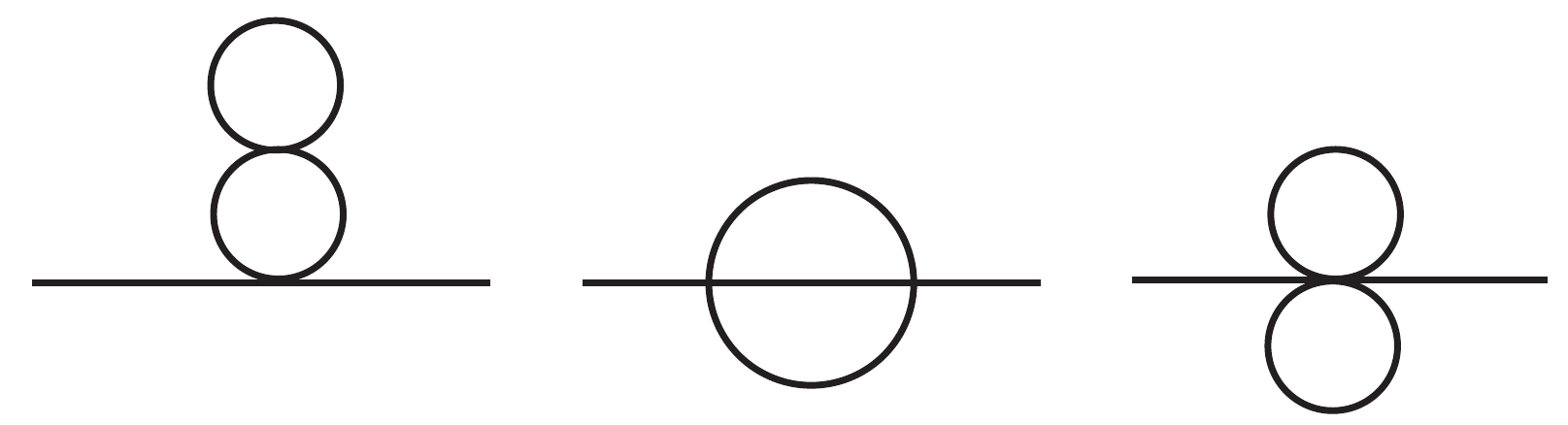}
\caption{The three two-loop topologies.}
\label{fig:2p}
\end{figure}
\bea 
\label{eq:NFSlimit}
\partial_\pm \rightarrow g^{\mp 1/2}\partial_\pm,\qquad \Theta_\pm\rightarrow g^{\mp 1/4}\Theta_\pm\,,
\eea 
where $\Theta_\pm=\frac{1}{2}(1\pm\Gamma_{11})\Theta$ are the left and right-moving components of the 32 component spinor in (\ref{action}). This is essentially a worldsheet dilation and boost and thus only affects the part of the Lagrangian which is not (worldsheet) Lorentz invariant. After light-cone and kappa symmetry gauge-fixing, the quadratic Lagrangian is completely Lorentz invariant and only a few terms in the quartic Lagrangian (\ref{eq:quarticL}) are non-invariant. Thus the limit (\ref{eq:NFSlimit}) will bring some terms of order $1/g$ in $\mathcal L_4$ to zeroth order, implying that already at leading order we have non-trivial interaction terms. Keeping only the leading order terms, the NFS Lagrangian becomes
\bea 
\label{eq:NFS}
\mathcal{L}^{nfs}=\mathcal L_2
+\gamma\Big\{
i\partial_-x^m\,\bar{\Theta}_-\Gamma^-\Gamma_m(x_1\Gamma^1-x_2\Gamma^2)\Theta_-
-(x_1^2-x_2^2)\,\big(\frac{1}{4}(\partial_- x_m)^2-i\bar{\Theta}_-\Gamma^-\partial_-\Theta_-\big)
\Big\}
\eea 
which is considerably simpler than (\ref{eq:quarticL}). Here we also performed a simple scaling of the worldsheet and fermionic coordinates resulting in an overall factor $\gamma$ in front of the quartic terms. This parameter essentially undoes the boost and is introduced for later convenience \cite{Klose:2007rz, Abbott:2011xp}. 

If we expand out the interaction terms involving fermions in terms of the eight coset fermions $\vartheta=\mathcal P_8\Theta$ and the 24 non-coset fermions $\upsilon=(1-\mathcal P_8)\Theta$, making use of the fact that $[\Gamma_1,\mathcal P_8]=[\Gamma_2,\mathcal P_8]=0$ and $\mathcal P_8\Gamma_{a'}\mathcal P_8=0$ where $a'=4,\ldots,9$ refers to the $T^6$-directions, we get
\bea 
&&i(x_1^2-x_2^2)\,\bar{\vartheta}_-\Gamma^-\partial_-\vartheta_-
-i(x_1\partial_-x^2+x_2\partial_-x^1)\,\bar{\vartheta}_-\Gamma^-\Gamma^{12}\vartheta_-
\nn\\
&&{}
+i(x_1^2-x_2^2)\,\bar{\upsilon}_-\Gamma^-\partial_-\upsilon_-
-i(x_1\partial_-x^2+x_2\partial_-x^1)\,\bar{\upsilon}_-\Gamma^-\Gamma^{12}\upsilon_-
\nn\\
&&{}
+2i\partial_-x^{a'}\,\bar{\upsilon}_-\Gamma^-\Gamma_{a'}(x_1\Gamma^1-x_2\Gamma^2)\vartheta_-
+i\partial_-x^{a'}\,\bar{\upsilon}_-\Gamma^-\Gamma_{a'}(x_1\Gamma^1-x_2\Gamma^2)\upsilon_-\,.
\label{eq:fermi-int}
\eea 
If instead of working with the full Green-Schwarz action for the string we would have started with the $AdS_2\times S^2$ supercoset sigma-model and added free massless bosons and fermions for the $T^6$ part we would only obtain the terms in the first line above. We will find that the other interaction terms, which involve also the non-coset fermions, are necessary in order to get the form of the dispersion relation that we expect from integrability. Therefore it appears that (despite its classical integrability) the supercoset model plus free fields does not seem to give the correct quantum corrections and the specific interaction terms present in the full GS string are crucial for quantum integrability.

\begin{figure}[t]
\centering
\includegraphics[scale=0.3]{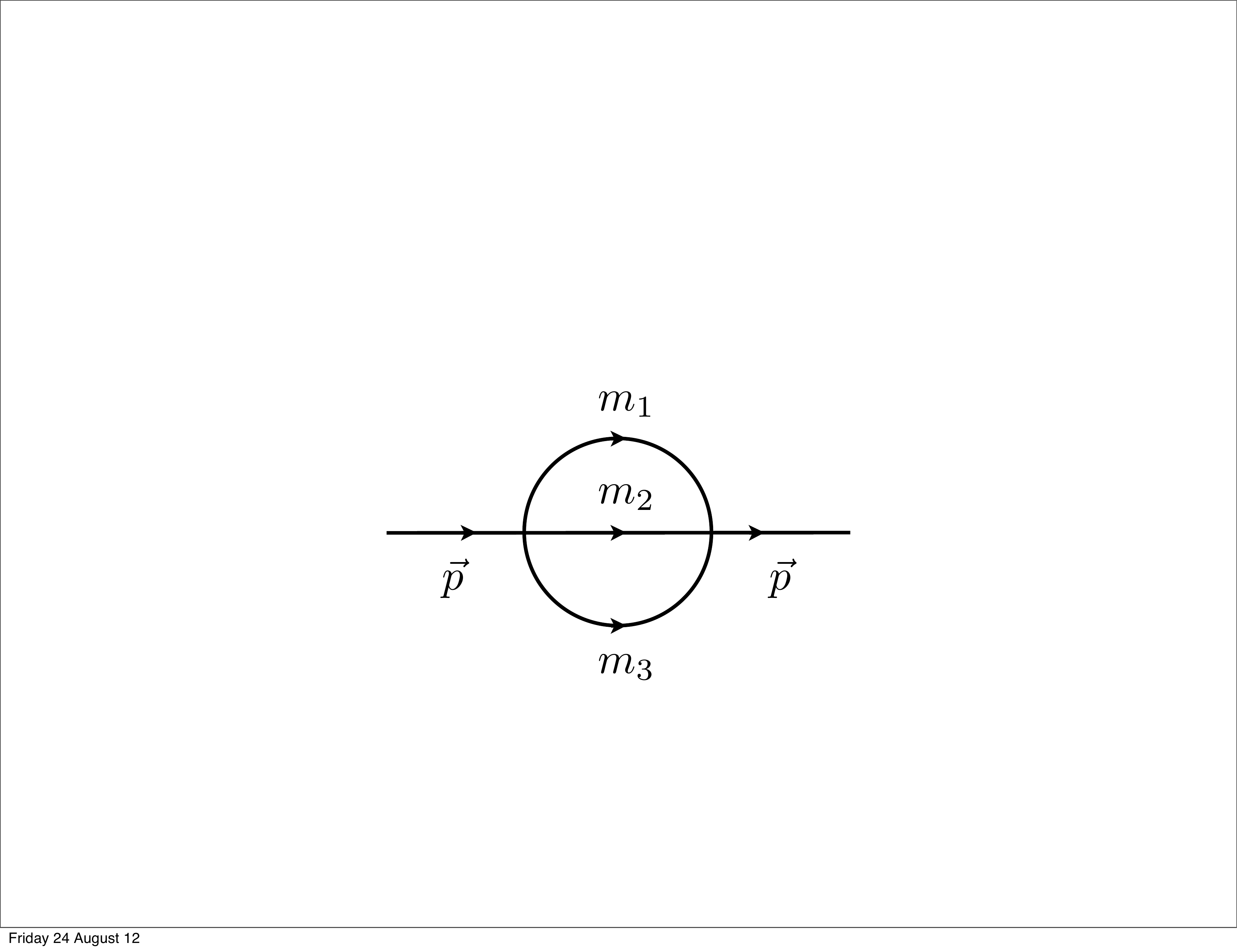}
\caption{The sunset / sunrise diagram with different internal masses.}
\label{fig:sunrise}
\end{figure}

Using (\ref{eq:NFS}) we will calculate the two-loop correction to the dispersion relation. The relevant diagrams are depicted in figure \ref{fig:2p} and generally one also needs the sixth order Lagrangian, but in the NFS limit these terms are subleading in the $1/g$ expansion and can be neglected to lowest order. The contributing four-vertex diagrams are a double tadpole and the sunset diagrams (also denoted sunrise or London transport diagram). For the first diagram we need the quartic fermion terms regardless of whether the external legs are bosonic or not. For the sunset diagram on the other hand, we only need the piece of the Lagrangian that is at most quadratic in fermions for external bosonic legs. Furthermore, the two-loop tadpole diagrams should vanish due to supersymmetry \cite{Klose:2007rz} which means that the only two-loop diagram we have to evaluate is the sunset diagram, as depicted in figure \ref{fig:sunrise}.
Thus, assuming supersymmetry we can calculate the two-loop correction to the propagator without having to know the $\mathcal{O}(\Theta^4)$ terms in the action.

The way to evaluate the diagram follows standard QFT technology with the special simplification of only having right-moving momenta in the loop integrals - a direct consequence of the NFS limit. Doing all the combinatorics using (\ref{eq:NFS}) results after some work in the following amplitude
\bea 
\label{eq:A-2loop}
&& i\mathcal{A}_{sun}=\frac{1}{2}\big(A_{040}+4A_{112}+4A_{121}+5A_{130}+4A_{211}+6A_{220}+5A_{310}
+A_{400}\big) \\ \nn 
&& +\frac{3}{2}\big(B^1_{022}+B^1_{112}+B^2_{112}+B^2_{202}+B^3_{130}+B^3_{310}\big)+3\big(C_{112}\big)\,,
\eea 
%
where we have defined
\bea \nn
A_{rst}=I_{rst}[1,1,1],\quad B_{rst}^1=I_{rst}[1,0,0],\quad B_{rst}^2=I_{rst}[0,1,0],\quad 
B^3_{rst}=I_{rst}[0,0,1],\quad
C_{rst}=I_{rst}[0,0,0]
\eea 
in terms of the standard sunset integral
\bea 
\label{eq:I}
&& I_{rst}[m_1,m_2,m_3]=\int \frac{d^2k d^2l}{(2\pi)^4}\frac{(k_-)^r(l_-)^s(p_--k_--l_-)^t}{(\vec k^2-m_1^2)(\vec l^2-m_2^2)((\vec p-\vec k -\vec l)^2-m_3^2)}\,,
\eea 
where $k$ and $l$ denotes the loop momenta. Introducing Feynman parameters to write
\bea \nn 
\frac{1}{ABC}=2\int_0^1 dx_1dx_2dx_3 \frac{\delta(x_1+x_2+x_3-1)}{\big(x_1 A+x_2 B+x_3 C\big)^3}
\eea 
and completing the squares in the denominators 
\bea 
l_\mu\rightarrow l_\mu+\frac{x_3}{x_2+x_3}(p_\mu-k_\mu), \qquad k_\mu\rightarrow k_\mu +\frac{x_2x_3}{x_1x_2+x_1x_3+x_2x_3}p_\mu 
\eea 
together with subsequent integration over $l$ and $k$ we end up with
\bea \nn 
&& I_{000}[m_1,m_2,m_3]=\int \frac{dx_1 dx_2 dx_3}{16\pi^2}\frac{\delta(x_1+x_2+x_3-1)}{(x_1x_2+x_1x_3+x_2x_3)\big(m_1^2x_1+m_2^2x_2+m_3^2x_3\big)-x_1x_2x_3\vec{p}^2}  \\ 
&& 
=\int \frac{dx_1 dx_2 dx_3}{16\pi^2} \rho[m_1,m_2,m_3]\,.
\eea 
This is the only integral we need, since any integral with powers of $l_-$ or $k_-$ in the numerator in (\ref{eq:I}) vanishes due to the fact that
\bea \nn
\int d^2k \frac{k_-^r}{(\vec k^2-\Delta)^m}=0,\qquad r,m>0\,.
\eea 
Performing the shifts and integrations in (\ref{eq:A-2loop}) we find
\bea \nn
i\mathcal{A}_{sun}=\frac{1}{16\pi^2}\int dx_1dx_2dx_3\Big(\gamma[1,1,1]+\gamma[1,0,0]+\gamma[0,1,0]+\gamma[0,0,1]+\gamma[0,0,0]\Big)p_-^4\,,
\eea 
where the notation indicates the diagrams with three, one and zero massive fields propagating in the loops. The $\gamma$'s are fairly complicated expressions involving the Feynman parameters explicitly given by
\bea \nn
&& \gamma[1,1,1]=\frac{x_3^2 (4 x_1^3 x_2^3 + 
   4 x_1^2 x_2^2 (x_1 + x_2) x_3 + (x_1^2 + x_1 x_2 + x_2^2) (x_1^2 + 4 x_1 x_2 + 
      x_2^2) x_3^2)}{2\big(x_1x_2+x_1x_3+x_2x_3\big)^4}\rho[1,1,1] \\ \nn
&& \gamma[1,0,0]=\frac{3 x_1^3 x_2^2 (x_1 + x_2) x_3^2}{2\big(x_1x_2+x_1x_3+x_2x_3\big)^4}\rho[1,0,0],\qquad 
\gamma[0,1,0]=\frac{3 x_1^2 x_2^3 (x_1 + x_2) x_3^2}{2\big(x_1x_2+x_1x_3+x_2x_3\big)^4}\rho[0,1,0]
\\ \nn 
&& \gamma[0,0,1]=\frac{3 x_1 x_2 (x_1^2 + x_2^2) x_3^4}{2\big(x_1x_2+x_1x_3+x_2x_3\big)^4}\rho[0,0,1],\qquad 
\gamma[0,0,0]=\frac{3 x_1^3 x_2^3 x_3^2}{\big(x_1x_2+x_1x_3+x_2x_3\big)^4}\rho[0,0,0]\,.
\eea 
Luckily Mathematica knows how to evaluate these integrals and in a hopefully obvious notation we get
\bea \label{eq:sunrise-result}
&& i\mathcal{A}_{sun}=\big(111\big)+\big(100\big)+\big(010\big)+\big(001\big)+\big(000\big)\\ \nn 
&&=\frac{1}{192\pi^2}\big(-3+\pi^2)p_-^4+\frac{5p_-^4}{384\pi^2}+\frac{5p_-^4}{384\pi^2}+\frac{p_-^4}{48\pi^2}-\frac{p_-^4}{32\pi^2}=\frac{\gamma^2}{192}p_-^4\,,
\eea 
where we've reinstated the NFS power counting parameter $\gamma$ in the last line, see (\ref{eq:NFS}). We see that we have a rather non-trivial cancellation between the odd looking $1/\pi^2$ terms coming from diagrams with different masses of internal fields. It is clear from this expression that if we had instead used only the supercoset $AdS_2\times S^2$ sigma model plus free massless fields, so that the only fermion interaction terms are those in the first line of (\ref{eq:fermi-int}), we would only get the first term above which involves only the massive fields. The $1/\pi^2$ terms would then not cancel and we would obtain a result which is difficult to reconcile with the one expected from integrability.

The computed two-loop correction shifts the bare pole of the propagator to
\bea \label{eq:2loop-pole}
\frac{i}{\vec p^2-1}\rightarrow \frac{i}{p_0^2-1-p_1^2+\frac{\gamma^2}{192}p_-^4}+\dots
\eea 
The next question we want to answer is, how does the above two-loop result fit into the exact dispersion relation? Expanding the dispersion relation from (\ref{eq:integrability-energy}) in $h(g)$ we find
\bea \nn
E=\sqrt{1+4h^2\sin^2\frac{p_1}{2}}=\sqrt{1+p_1^2-\frac{1}{192h^2}p_-^4}+\dots
\eea 
where we used that $p_1=\frac{1}{2}(p_+-p_-)$ in the last line and kept only the leading $p_-^4$ piece. Intriguingly the above is exactly what we find in (\ref{eq:2loop-pole}) if we identify
\bea 
\gamma=\frac{1}{h}\,.
\eea 
Thus we see that the sine structure of the dispersion relation holds also in the \2 background. Furthermore, we have seen that this result depends on the massless modes and specifically on the precise structure of their interactions in the GS Lagrangian. Removing these interaction terms would give a result inconsistent with the sine square expression expected from integrability. We regard this a rather strong evidence for the quantum integrability of the \2 GS string.

\section{Conclusion}
We have investigated both classical and quantum aspects of the type IIA \2 string in a BMN expansion. String theory on this background is interesting for many reasons. For example, it has a four-dimensional sector which arises in many black hole physics applications. Furthermore, and perhaps more closely related to the analysis of this paper, the curved part can be realized as a supercoset model and in fact the full GS action is classically integrable \cite{Sorokin:2011rr,Cagnazzo:2011at}. This makes the powerful tools developed in the context of AdS / CFT integrability accessible. It seems clear, however, that these tools will need to be generalized somewhat in order to deal with the massless modes which play an important role in the integrability.

In this paper we have investigated classical energies and loop corrected propagators. Our findings fit nicely into the general framework of integrability. Our analysis began with a computation of simple energy corrections to bosonic string states in the transverse part of $AdS_2\times S^2$. These we then compared to the predictions of a set of quantum Bethe equations conjectured in \cite{Sorokin:2011rr}. The Bethe equations are defined through a set of integral equations which arise in the classical / thermodynamic limit. These integral equations, in turn, are expressed in terms of the components of a flat current that arises in the coset formulation. We found that the prediction of the Bethe equations agreed precisely with the computed string energies. 

We also computed one and two-loop contributions to two-point functions. These are constructed from the string coordinates which can be thought of as fields in a (1+1)-dimensional (worldsheet) QFT. Integrability dictates that the loop corrections to the propagators should fit into a (non-relativistic) dispersion relation. Based on prior work in $AdS_5\times S^5$ and $AdS_3\times S^3\times T^4$ it is expected that the one-loop correction should be zero and this was indeed what we found. While the one-loop correction to the massless modes is trivially zero, the loop corrections for the massive excitations are zero due to a delicate cancellation between boson and fermion loops. Thus the first non-trivial correction to the dispersion relation enters at the two-loop level. While a two-loop computation generally demands the full quartic and sixth-order Lagrangian, which is not at present available, it is nevertheless possible to compute the two-loop correction in the so-called near flat space (NFS) limit. In this limit the right-moving worldsheet sector is boosted which results in a vastly simplified theory. What is more, the NFS Lagrangian has no explicit coupling dependence, thus the two-loop contribution to the NFS expanded dispersion relation arises solely from a two-loop four-vertex sunset diagram. By explicit computation we found that the two-loop contribution fits nicely into the sine-squared dispersion relation, analogous to the NFS $AdS_5 \times S^5$ string \cite{Klose:2007rz}. This fact was seen to depend strongly on using the complete GS string action, if the action is truncated to the supercoset sector (together with free massless fields) one obtains the wrong result. This gives a strong indication that the GS string action is integrable at the quantum level. Unlike in other examples of AdS/CFT there is no kappa symmetry gauge-fixing of the \2 GS string which gives the supercoset model (the supercoset modes has only 8 of the required 16 fermions).

The $AdS_2/CFT_1$ duality remains largely unexplored and there are several possible continuations of this work. For example, it would be interesting to investigate how the massless and massive modes interact in more detail. The natural aim for this investigation is to learn how to incorporate the massless modes as excitations in the Bethe ansatz solution. A stepping stone for this analysis would be to compute the $2\rightarrow 2$ S-matrix \cite{Staudacher:2004tk,Beisert:2005tm}. This is currently under investigation. Understanding what the dual $CFT_1$ is would also be of great interest.

Furthermore, it would also be interesting to investigate various spinning and folded string configurations, basically continuing the research initiated in \cite{Sorokin:2011rr}. This would yield more information on the fairly unknown quantum sector of the theory. We leave this for future work. 

\section*{Acknowledgments}
It is a pleasure to thank M. Abbott, T. McLoughlin and N. Rughoonauth for valuable discussions. JM acknowledges support from the National Research Foundation (NRF) of South Africa under the Thuthuka and Incentive Funding for Rated Researches programmes.  PS is supported by a postdoctoral grant from the Claude Leon Foundation. The research of LW is supported in part by NSF grants PHY-0555575 and PHY-0906222.

\appendix

\section{Bosonic parameterization}
We write the metric of \2 in terms of global coordinates following closely \cite{Rughoonauth:2012qd}
\bea 
ds^2=-\Big(\frac{1+\frac{1}{4}x_1^2}{1-\frac{1}{4}x_1^2}\Big)^2dt^2+\frac{dx_1^2}{\big({1-\frac{1}{4}x_1^2}\big)^2}+
\Big(\frac{1-\frac{1}{4}x_2^2}{1+\frac{1}{4}x_2^2}\Big)^2d\varphi^2+\frac{dx_2^2}{\big({1+\frac{1}{4}x_2^2}\big)^2}+dx_{a'}^2\,,
\eea 
where $a'=3,\dots,8$ denotes the $T^6$ directions. The spin connection of the background is readily calculated from the vanishing of the torsion,
\bea \nn 
de^A+e^B \omega_B{}^A=0
\eea 
which gives
\bea 
\label{eq:spin-connection-rep}
\omega^{01}=-\frac{x_1 dt}{1-\frac{1}{4}x_1^2},\qquad \omega^{23}=-\frac{x_2 d\varphi}{1+\frac{1}{4}x_2^2}\,.
\eea 

\section{$\Theta$ in terms of real variables}
When needed we use the representation of $\Gamma$-matrices and the charge conjugation matrix $\mathcal{C}$ given in \cite{Babichenko:2009dk}.
We can pick a real representation $\chi_\pm^c$ for the worldsheet fermions which diagonalizes the quadratic BMN action. In terms of the original 32-component Majorana spinor $\Theta$ we have
\bea 
\nn
\Theta=\frac{1}{4\sqrt{2}}\Theta_1\oplus\Theta_2\oplus\Theta_3\oplus\Theta_4\,,
\eea 
where the 8-component spinors $\Theta_i$ are given by
\begin{small}
\bea \label{eq:theta}
&& \Theta_1=
\left(\begin{array}{c}
i \chi_+^5+\chi_+^6-i \chi_+^7-\chi_+^8 \\ 
-\chi_+^5-i \chi_+^6-\chi_+^7-i \chi_+^8 \\
i \chi_+^5+\chi_+^6+i \chi_+^7+\chi_+^8 \\ 
\chi_+^5+i \chi_+^6-\chi_+^7-i \chi_+^8 \\ 
\chi_+^5+i \chi_+^6-\chi_+^7-i \chi_+^8 \\ 
-i \chi_+^5+\chi_+^6+i \chi_+^7-\chi_+^8 \\
\chi_+^5-i \chi_+^6-\chi_+^7+i \chi_+^8 \\
i \chi_+^5-\chi_+^6+i \chi_+^7-\chi_+^8
\end{array} 
\right),\qquad 
\Theta_2=
\left(\begin{array}{c}
-i \chi_+^1-\chi_+^2+i \chi_+^3+\chi_+^4 \\
\chi_+^1+i \chi_+^2+\chi_+^3+i \chi_+^4 \\
-i \chi_+^1-\chi_+^2-i \chi_+^3-\chi_+^4 \\
-\chi_+^1-i \chi_+^2+\chi_+^3+i \chi_+^4 \\
\chi_+^1-i \chi_+^2+\chi_+^3-i \chi_+^4 \\
-i \chi_+^1+\chi_+^2+i \chi_+^3-\chi_+^4 \\
\chi_+^1-i \chi_+^2-\chi_+^3+i \chi_+^4 \\
i \chi_+^1-\chi_+^2+i \chi_+^3-\chi_+^4
\end{array} 
\right) \\ \nn 
&& \Theta_3=
\left(\begin{array}{c}
i \chi_-^1+\chi_-^2-i \chi_-^3+\chi_-^4 \\
\chi_-^1+i \chi_-^2+\chi_-^3-i \chi_-^4 \\
i \chi_-^1+\chi_-^2+i \chi_-^3-\chi_-^4 \\
-\chi_-^1-i \chi_-^2+\chi_-^3-i \chi_-^4 \\
\chi_-^1-i \chi_-^2+\chi_-^3+i \chi_-^4\\
i \chi_-^1-\chi_-^2-i \chi_-^3-\chi_-^4 \\
\chi_-^1-i \chi_-^2-\chi_-^3-i \chi_-^4 \\
-i \chi_-^1+\chi_-^2-i \chi_-^3-\chi_-^4
\end{array} 
\right),\qquad 
\Theta_4=
\left(\begin{array}{c}
-i \chi_-^5+\chi_-^6-i \chi_-^7+\chi_-^8 \\
-\chi_-^5+i \chi_-^6+\chi_-^7-i \chi_-^8 \\
-i \chi_-^5+\chi_-^6+i \chi_-^7-\chi_-^8 \\
\chi_-^5-i \chi_-^6+\chi_-^7-i \chi_-^8 \\
\chi_-^5+i \chi_-^6-\chi_-^7-i \chi_-^8 \\
i \chi_-^5+\chi_-^6+i \chi_-^7+\chi_-^8 \\
\chi_-^5+i \chi_-^6+\chi_-^7+i \chi_-^8 \\
-i \chi_-^5-\chi_-^6+i \chi_-^7+\chi_-^8
\end{array} 
\right)\,.
\eea 
\end{small}

\end{document}